\documentclass[pra,aps,10pt,twocolumn,showpacs,superscriptaddress]{revtex4-2}
\usepackage{amsmath,amssymb}
\usepackage{xcolor}
\usepackage[allcolors=blue,colorlinks=true]{hyperref}
\usepackage{bm}
\usepackage{graphicx}

\usepackage{mathptmx}
\DeclareMathAlphabet{\mathcal}{OMS}{cmsy}{m}{n}

\newcommand{\llangle}{\langle\!\langle}
\newcommand{\rrangle}{\rangle\!\rangle}
\renewcommand{\Im}{\mathop{\mathrm{Im}}}

\begin{document}

\title{Topological charges and parity selection at Floquet quasienergy degeneracies}

\author{Sigmund Kohler}
\affiliation{Quantum Advanced Research Center, CSIC, 28049 Madrid, Spain}
\affiliation{Instituto de Ciencia de Materiales de Madrid, CSIC, 28049 Madrid, Spain}

\author{David Gu\'ery-Odelin}
\affiliation{Laboratoire Collisions Agr\'egats R\'eactivit\'e, UMR 5589,
FERMI, Universit\'e de Toulouse, CNRS, 118 Route de Narbonne, 31062 Toulouse CEDEX 09, France}

\date{\today}

\begin{abstract}
The quasienergy spectrum of a strongly driven two-level system as a
function of the driving parameters exhibits conical intersections, which
are enabled by hidden time-nonlocal symmetries. We show that each such
crossing carries a quantized topological charge: the Floquet--Berry phase
acquired along an adiabatic loop around a cone is equal to a
$\mathbb{Z}_2$-valued charge.  We further identify a second family of
degeneracies that occurs at vanishing driving amplitude, when the level
splitting matches $m$ energy quanta of the field.  Along the Stark-shifted
resonance line, the minimum quasienergy gap opens as $|A|^m$, and the
charge is nontrivial only for odd~$m$. We analytically derive both results
from a perturbative reduction to a spin-$1/2$ in an effective
two-dimensional magnetic field and confirm them numerically through the
Bargmann invariant. Moreover, we propose a chirality-based protocol that
cancels the dynamical phase to isolate the geometric one, and an
ancilla-based Ramsey readout that renders the topological charge directly
observable.
\end{abstract}

\maketitle

\section{Introduction}

Geometric phases organize the response of quantum systems to slow parameter
variation \cite{XiaoRMP10}. For a two-level system in a slowly varying
magnetic field, the Berry phase acquired along a loop equals half the solid
angle swept by the field, and in two dimensions a loop enclosing a generic
degeneracy contributes $\pi$ \cite{BerryPRSA84b,RestaJP00,XiaoRMP10}. The
degeneracy acts as a topological monopole, its double cone (or diabolic
point) marking its locus in parameter space.

Do such structures survive when the two-level system is strongly driven? For
weak resonant driving, a rotating-wave approximation maps the problem back
onto a static spin-$1/2$ in an effective magnetic field
\cite{LeekS07,WeinbergPR17}, recovering the standard Berry monopole. Far
from resonance and at large amplitude it fails, and the full Floquet
formalism applies \cite{ShirleyPR65,SambePRA73}: the adiabatic
invariants are Floquet modes rather than instantaneous eigenstates
\cite{AharonovPRL87,MooreJPA90b}, whose slow transport is governed by a
Berry connection in Sambe space \cite{WeinbergPR17}.

The quasienergy spectrum in this regime has been studied extensively,
starting from coherent destruction of tunneling \cite{GrossmannPRL91},
rooted in the dynamic localization of driven charges
\cite{DunlapPRB86,HolthausPRL92}, where exact crossings on the amplitude
axis are enabled by a generalized parity \cite{PeresPRL91}. They have been
observed with Bose--Einstein condensates in shaken optical lattices
\cite{LignierPRL07,ArnalPRA20}, optical waveguides \cite{DellaVallePRL07},
double quantum dots \cite{StehlikPRB12,ForsterPRL14}, and superconducting
qubits \cite{SillanpaaPRL06,BernsNL08}. Exact crossings were recently shown
to occur also when the detuning is an integer multiple of the driving
frequency \cite{KohlerQ26}; a static detuning breaking this symmetry turns
each into a conical intersection, around which the open-system dissipative
response develops a characteristic structure \cite{KohlerPRA24,KohlerJCP25}.

Whether these conical intersections carry a topological charge in the sense
of the Berry phase has not, to our knowledge, been addressed. The question
is relevant for three reasons. First, the connection lives in Sambe space,
the product of Hilbert space with $T$-periodic functions
\cite{SambePRA73,WeinbergPR17}, so the Floquet modes depend on time both
explicitly and implicitly through their slow parameter variation.
Second, the dynamical phase generically dominates the
geometric one, requiring a compensation scheme. Third, a distinct,
non-conical family of degeneracies arises at $m$-photon resonances and zero
amplitude, whose topological status is unclear.

The Berry phase of a driven
spin was measured in superconducting qubits within the rotating-wave
approximation \cite{LeekS07}. Quantized geometric responses appear in
quasi-periodically driven two-level systems, where two incommensurate drives
realize a synthetic Brillouin zone with Floquet--Dirac
points \cite{MartinPRX17,BoyersPRL20}, and the Berry curvature of a
Floquet--Bloch band was reconstructed in a driven optical lattice
\cite{FlaschnerS16}. Floquet--Dirac points in dual-dressed qubits were
identified spectroscopically, with the Berry-phase connection left open
\cite{FregosiSR23}, while high-frequency approximations map time-dependent
lattices to static ones with nontrivial topology
\cite{KitagawaPRB10,GomezLeonPRL13,EckardtRMP17,PerezGonzalezPRL19}. None addresses the topology of exact crossings of a monochromatically
driven two-level system at strong driving, outside a rotating-wave or
high-frequency approximation.

A simple picture anticipates this charge. Time-reversal symmetry renders the
Floquet modes real in Sambe space, so adiabatic transport defines a real
line bundle over the parameter plane. A real normalized state can return
only up to a sign after a closed loop, so the Floquet--Berry phase is
restricted to $0$ and $\pi$, recording the sign it returns with. Where the modes stay real the curvature vanishes, and the topology
concentrates in pointlike $\mathbb{Z}_2$ vortices at the degeneracies, where
the real eigenline ceases to exist. The charge we
study is the winding of this real bundle, $\mathbb{Z}_2$-valued by
construction.

We establish three results. First, every conical intersection at integer
detuning carries a nontrivial $\mathbb{Z}_2$ charge, detected by a
Floquet--Berry phase $\gamma = \pi \pmod{2\pi}$ and derived analytically,
without a high-frequency approximation, from a mapping to a spin-$1/2$ in a
constant two-dimensional field. Second, the degeneracies at an $m$-photon resonance and zero
amplitude obey a parity selection rule: the charge is $\pi$ for odd $m$ and
trivial for even $m$. Third, we propose a protocol that exploits the
chirality of the Floquet Hamiltonian to cancel the dynamical phase over two
symmetry-related halves of a loop, leaving the charge as the measurable
interference phase; unlike the standard spin-echo of the
rotating-wave regime, it is specific to the Floquet setting.

The paper is organized as follows. Section~\ref{sec:framework} sets up the
Floquet--Berry phase in Sambe space and the model Hamiltonian.
Section~\ref{sec:topology} establishes the two families of degeneracies and
their topological charges. Section~\ref{sec:protocol} presents the
chirality-based dynamical-phase cancellation, which in
Sec.~\ref{sec:measurement} is extended to an ancilla-based measurement
protocol. Section~\ref{sec:conclusion} concludes, while details of the
perturbation theory and of symmetry relations can be found in the Appendix.

\section{Floquet--Berry phase and model}
\label{sec:framework}

We consider a Hamiltonian $H(\bm{x},t) = H(\bm{x},t+T)$ that depends
periodically on time with period $T = 2\pi/\Omega$ and parametrically on
a set of slow variables $\bm{x}$. For fixed $\bm{x}$, the Schr\"odinger
equation admits Floquet solutions \cite{ShirleyPR65,SambePRA73}
\begin{equation}
|\psi(t)\rangle = e^{-iq(\bm{x})t}\,|\phi(\bm{x},t)\rangle ,
\end{equation}
with the quasienergy $q(\bm x)$ and the $T$-periodic Floquet mode 
$|\phi(\bm{x},t)\rangle$ obeying the eigenvalue equation (in units with $\hbar=1$)
\begin{equation}
\big(H(\bm{x},t)-i\partial_t\big)|\phi(\bm{x},t)\rangle 
= q(\bm{x})|\phi(\bm{x},t)\rangle .
\label{eq:Feq}
\end{equation}
For slow cyclic variation of $\bm{x}$ along a loop $\mathcal{C}$ with
period $T^*=NT$ and integer $N\gg 1$, adiabatic following to the Floquet 
mode yields the Floquet--Berry phase \cite{WeinbergPR17}
\begin{equation}
\gamma = i\oint_{\mathcal{C}} d\bm{x}\cdot\llangle\phi(\bm{x})|\partial_{\bm{x}}
\phi(\bm{x})\rrangle ,
\label{eq:FBloop}
\end{equation}
where the double bracket denotes the inner product in Sambe space
(Hilbert space tensored with $T$-periodic functions, with Dirac notation
$|\phi(t)\rangle \equiv \langle t|\phi\rrangle$ \cite{SambePRA73}), equivalently the
time average over one driving period. Equation~\eqref{eq:FBloop} is the
Floquet analog of the Berry phase, with the static eigenstate replaced
by a Floquet mode. 
If a smooth Floquet gauge exists on a surface $\mathcal S$ bounded by
$\mathcal C$, while $\mathcal S$ contains no degeneracy, Stokes' theorem gives
$\gamma = \int_{\mathcal S} dx_1\,dx_2\,F_{12}$, with
$F_{12}=-2\Im\llangle\partial_1\phi|\partial_2\phi\rrangle$
\cite{GargAJP10,RestaJP00,XiaoRMP10}. In the present time-reversal-symmetric
problem $F_{\epsilon A}=0$ away from degeneracies, yet a loop enclosing a
degeneracy may still carry $\gamma=\pi$: a globally smooth single-valued
real Floquet gauge then does not exist on a spanning surface containing the
singularity. The resulting $\mathbb Z_2$ phase is equivalently the holonomy
of the real Floquet eigenline bundle.

\begin{figure}
\centerline{\includegraphics{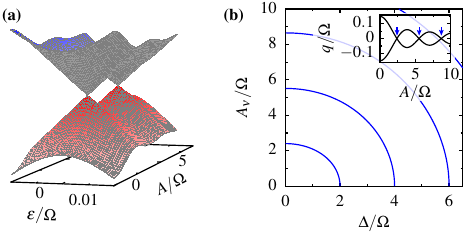}}
\caption{Floquet spectrum of the driven two-level system.
(a) Quasienergies as a function of the amplitude $A$ and the detuning
$\epsilon$ for tunnel coupling $\Delta=0.3\Omega$.
(b) Location of the $\nu$th exact crossing as a function of the tunneling
for $\epsilon=0$.
Inset: Corresponding quasienergies, where the arrows mark the crossings shown
in the main panel.}
\label{fig:spectrum}
\end{figure}

We specialize to the driven two-level system
\begin{equation}
H(t) = \frac{\Delta}{2}\sigma_x + \frac{1}{2}\big(\epsilon + A\cos\Omega t\big)\sigma_z ,
\label{eq:H}
\end{equation}
with slow parameters $\bm{x}=(\epsilon,A)$ and fixed $\Delta$, $\Omega$,
for which the Floquet Hamiltonian $\mathcal{H} = H(t)-i\partial_t$
possesses three relevant symmetries. For $\epsilon=0$, the generalized
parity $\sigma_x \otimes (t\to t+T/2)$ enables exact quasienergy crossings
on the $A$-axis \cite{PeresPRL91} visualized in Fig.~\ref{fig:spectrum}.
Time-reversal symmetry, $t\to -t$ combined with complex
conjugation, keeps the Sambe-space components real and
makes the Berry curvature $F_{\epsilon A}$ vanish away from degeneracies.
Finally, a chirality $S = \sigma_y\otimes(t\to -t)$ satisfying
$S\mathcal{H}S^{-1}= -\mathcal{H}$ pairs Floquet modes of opposite
quasienergy \cite{EngelhardtPRL21, KohlerQ26}.
In addition, a hidden time-nonlocal symmetry at integer detuning
$\epsilon=n\Omega$ generates further exact crossings \cite{KohlerQ26}.

Throughout, the amplitude $A$ is understood as a signed parameter.
Although the frozen Floquet spectra at $A$ and $-A$ are related by the
half-period time translation $-A\cos\Omega t = A\cos(\Omega t+\pi)$, the two signs
represent distinct points of the control manifold once the drive phase is
referenced to a fixed laboratory clock. A smooth sign change is implemented by
taking the real drive quadrature continuously through zero, so that the line
$A=0$ is an ordinary line of the $(\epsilon,A)$ plane that a loop may enclose.
This is what makes the degeneracies at vanishing amplitude
(Sec.~\ref{sec:topology}) encircleable. The resulting charge is a property of the
control loop and does not depend on the arbitrary choice of Floquet gauge, since
the Bargmann phase~\eqref{eq:bargmann} is invariant under the local $U(1)$
redefinition of each Floquet mode. Its physical, gauge-independent nature
is borne out by the interferometric protocol of Sec.~\ref{sec:measurement},
which detects the odd-$m$ charge at vanishing amplitude directly.

\section{Topological charges of the degeneracies}
\label{sec:topology}

In the presence of time-reversal symmetry, according to the Wigner--von Neumann
theorem, two system parameters must be adjusted to find an accidental
degeneracy of a hermitian operator \cite{Haake2018}. Therefore, we expect
to find them at isolated points in the $(\epsilon,A)$ plane.  A well-known
example are the exact crossings for $\epsilon=0$ and particular values of
$A$, which for small tunneling are determined by the zeros of a Bessel
function \cite{GrossmannEL92}.  With increasing tunneling, they shift to
smaller amplitudes \cite{CreffieldPRB03,DellaVallePRL07,WeiJCP26}, see
Fig.~\ref{fig:spectrum}(b).  In our model, a second family of crossings
occurs at $m$-photon resonances.  We analyze each by perturbation theory in
the Sambe-space neighborhood of the degeneracy.  Details of the
calculations can be found in Appendix~\ref{app:perturbation}.

\subsection{Degeneracies at integer detuning}

For a detuning $\epsilon=n\Omega$ with integer $n$, a hidden symmetry enables
crossings for certain amplitudes, while the corresponding Floquet modes can
be classified as even and odd \cite{KohlerQ26}. We consider such a crossing
with location generically denoted by $\epsilon_0$ and $A_0$. The Floquet
Hamiltonian is conveniently decomposed as $\mathcal{H}=\mathcal{H}_0+V+W$
with $V=(\epsilon-\epsilon_0)\sigma_z/2$ and $W=(A-A_0)\sigma_z\cos(\Omega
t)/2$.  Since we are interested in the behavior arbitrarily close to
$(\epsilon_0,A_0)$, it is sufficient to consider a two-dimensional subspace
of Sambe space, i.e., to use two degenerate Floquet modes of $\mathcal{H}_0$
as a basis.  We denote such a pair of degenerate states by
$|\varphi_\pm\rrangle$, where the index refers to the hidden parity.  Using
them as basis, yields the effective Floquet Hamiltonian
\begin{equation}
\mathcal{H}_\text{eff}
= (\epsilon-\epsilon_0)\,\vec v\cdot\vec\tau
  + (A-A_0)\,\vec w\cdot\vec\tau ,
\label{eq:Heff-integer}
\end{equation}
see Appendix~\ref{app:perturbation}.
The notation $\vec\tau$ for the vector of Pauli matrices emphasizes the
change of basis.  Due to time-reversal symmetry, the vectors $\vec v$,
$\vec w$ lie in the $xz$-plane.  This maps the Floquet problem near the
degeneracy to a spin-$1/2$ in a two-dimensional magnetic field $\vec
B(\epsilon,A) = (\epsilon-\epsilon_0) \vec v + (A-A_0)\vec w$.  For a
genuine conical intersection the Jacobian has rank two, equivalently $(\vec
v\times\vec w)_y\neq0$. If this condition fails, the touching is of higher
order and requires a separate analysis.  Provided $\vec v\nparallel\vec w$,
the local chirality $\chi = \operatorname{sgn}\det(\vec v,\vec w)=\pm1$
fixes the orientation with which $\vec B$ winds around the origin. The
Floquet--Berry phase depends on this winding only modulo $2\pi$, and is
therefore blind to the value of $\chi$: opposite chiralities carry the same
$\mathbb{Z}_2$ charge.  Then a small loop in parameter space around
$(\epsilon_0,A_0)$ winds the effective $\vec B$ exactly once around the
origin, and the corresponding Berry phase is $\pi \pmod{2\pi}$. Each
conical intersection at integer detuning therefore carries a nontrivial
$\mathbb{Z}_2$ charge; a loop enclosing $N_{\mathcal C}$ of them yields
$\gamma(\mathcal{C}) = \pi N_{\mathcal C}\ (\mathrm{mod}\ 2\pi)$.

\subsection{Degeneracies at vanishing amplitude}

A second family arises at $A_0 = 0$. There, the Hamiltonian is
time-independent with eigenstates $|g\rangle$, $|e\rangle$ and energies
$E_{g,e}$. Formally, these states are Floquet modes with quasienergies $q_-=E_g$
and $q_+= E_e$. When the energy splitting matches a multiple of the driving
frequency,
\begin{equation}
E_e - E_g = m\Omega ,
\label{eq:resonance}
\end{equation}
with integer $m$, it is convenient to replace $|e\rangle$ by the equivalent
Floquet mode $|\varphi_+(t)\rangle = e^{-im\Omega t}|e\rangle$.  Together
with $|\varphi_-(t)\rangle = |g\rangle$, it can be used as a starting point
for a degenerate perturbation theory.

In this basis the perturbation $V$ is diagonal and behaves
like in the case of integer detuning.  By contrast, the perturbation $W$
behaves qualitatively differently. In Sambe space, its off-diagonal matrix
elements $\llangle\varphi_-|\ldots|\varphi_+\rrangle$ involve the time
integral $\int_0^T dt\, e^{im\Omega t}\cos\Omega t$, which vanishes for
$m\neq 1$.  The lowest non-vanishing contribution arises at order $W^m$ and
is directed along $\tau_x$, while time-reversal symmetry forbids a term with $\tau_y$

For consistency, we also have to consider terms up to order $V^m$. Then the
effective two-level Hamiltonian in the degenerate subspace has
the generic form
\begin{equation}
\mathcal{H}_{\rm eff} = d_x\tau_x + d_z\tau_z ,
\label{eq:Wzero}
\end{equation}
with the coefficients
\begin{align}
d_x &= \beta_m A^m ,
\label{eq:dx}\\
d_z &= \alpha(\epsilon-\epsilon_0) + \delta_{\rm Stark}(\epsilon,A) ,
\label{eq:dz}
\end{align}
following from a Sambe-space perturbation expansion, while there are no
terms proportional to the identity or $\sigma_y$, as is justified in
Appendix~\ref{app:perturbation}.
The coefficient $\beta_m$ is the amplitude of the $m$ virtual one-photon transitions,
which vanishes only on non-generic parameter manifolds.
The last term in Eq.~\eqref{eq:dz} is an ac Stark shift, which does not depend on the driving phase.
Therefore, it must be an even
function of the amplitude, $\delta_{\rm Stark}(\epsilon,A) = \delta_{\rm Stark}(\epsilon,-A)$.  It
displaces the resonance line locally, but does not affect the topology,
which is controlled entirely by $d_x$.  For $m>1$ the degeneracy is no
longer diabolic: the dispersion is linear in $\epsilon$ while, along the
Stark-shifted resonance line $d_z=0$, the minimum gap opens as $|A|^m$, set by
the transverse coupling $d_x=\beta_m A^m$.

The topological charge follows from the winding of the planar field $\vec
B(\epsilon,A) = \big(d_x,\,0,\,d_z\big)$ of Eq.~\eqref{eq:Wzero} along a loop
encircling the degeneracy and crossing $A=0$ on both sides. There,
$d_x=\beta_m A^m$ reverses sign only for odd $m$, whereas $d_z$---being
even in $A$ up to the linear term $\alpha(\epsilon-\epsilon_0)$---returns to
itself. Therefore, the Stark corrections merely deform the contour in the $d_z$
direction. Hence $\vec B$ winds exactly once around the origin for odd $m$
and not at all for even $m$. We therefore obtain the parity selection rule
\begin{equation}
\gamma\big|_{\text{loop around }m\text{-photon resonance at }A=0}
= \begin{cases} \pi & \text{$m$ odd} ,\\ 0 & \text{$m$ even} . \end{cases}
\label{eq:parity}
\end{equation}
The odd-$m$ case is a genuine topological charge despite the absence of a
diabolic structure; the even-$m$ case is a degeneracy with trivial topology.
To our knowledge, this parity selection rule has not been
reported in the literature on driven two-level systems.
Such multiphoton power-law gap openings are themselves familiar in
Floquet perturbation theory~\cite{HolderPRA05}; what is specific here is
that the parity of this order determines a $\mathbb{Z}_2$ topological charge.

\subsection{Numerical verification}

We compute the Floquet--Berry phase \eqref{eq:FBloop} for small circular
loops of radius $r$ centered at arbitrary $(\epsilon,A)$ via the Bargmann
invariant \cite{BargmannJMP64}. In doing so, we discretize the loop and at
each point numerically solve the Floquet equation \eqref{eq:Feq}. 
The loop integral is computed in a standard way \cite{Asboth2016} by the overlaps of
Floquet modes in Sambe space at neighboring grid points along the loop.
Explicitly, the loop is discretized into points
$\bm{x}_0,\bm{x}_1,\dots,\bm{x}_L=\bm{x}_0$; following the Floquet mode
$|\phi(\bm{x})\rrangle$ by maximal Sambe overlap from point to point, one
forms the unnormalized Bargmann product
\begin{equation}
Z = \prod_{j=0}^{L-1}
\llangle\phi(\bm{x}_j)|\phi(\bm{x}_{j+1})\rrangle ,
\label{eq:bargmann}
\end{equation}
which is manifestly invariant under the local $U(1)$ gauge of each mode and whose
phase is the Floquet--Berry phase $\gamma=-\arg Z$.  Its modulus defines the
Bargmann discretization fidelity $\mathcal{F}_{\rm B}=|Z|^2$,
which tends to unity in the continuum limit and drops below it for a finite grid
when the loop passes close to a degeneracy. In a continuum limit, the
normalized product $Z/|Z|$ converges to $e^{-i\gamma}$ \cite{SimonPRL93,Asboth2016,RestaJP00}.

\begin{figure}
\includegraphics{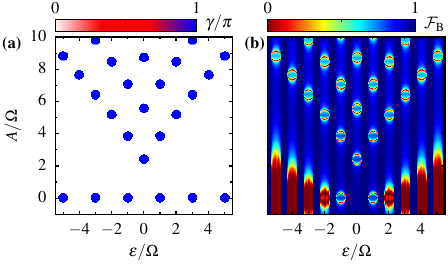}
\caption{(a) Floquet--Berry phase of the Floquet mode followed by
Sambe overlap along the loop, $Z$ [Eq.~\eqref{eq:bargmann}],
for an ellipse around $(\epsilon,A)$ with axes $r_\epsilon = r_A = 0.3\Omega$.
The tunnel matrix element is $\Delta=\Omega/10$.
(b) Bargmann discretization fidelity $\mathcal{F}_{\rm B}=|Z|^2$
for 500 grid points.}
\label{fig:bargmann} 
\end{figure}

As can be seen in Fig.~\ref{fig:bargmann}(a), the numerical results confirm
$\gamma=\pi \pmod{2\pi}$ at the integer-detuning crossings and at the odd-\(m\) zero-amplitude resonances, while the even-\(m\) resonances remain topologically trivial. Time-reversal symmetry ensures that $Z$ is real,
such that $\gamma$ can assume only values that are multiples of $\pi$.
Since $\gamma$ is extracted from instantaneous Floquet eigenmodes evaluated
along the contour, and not from a dynamical trajectory, its quantized value
is fairly robust against discretization errors and local gauge choices. By
contrast, the discretization fidelity $\mathcal{F}_{\rm B}$ depicted in
Fig.~\ref{fig:bargmann}(b) drops markedly near degeneracies, more so for
the broader $m$-photon resonances at $A=0$ which become harder to resolve
on a finite grid as $m$ grows.

\section{Cancelling the dynamical phase}
\label{sec:protocol}

The total phase accumulated during a cyclic evolution contains, in
addition to the geometric contribution $\gamma$, a dynamical part 
$\Phi = -\int_0^{T^*} q(\bm{x}(t))\,dt$ which is generally much larger and varies
sensitively with the loop. Isolating $\gamma$ requires a compensation
scheme. The spin-echo trick of traversing the loop twice with a
$\pi$-pulse in between, used in the rotating-wave regime
\cite{LeekS07}, is not suitable here: it removes the dynamical phase at the
price of doubling the geometric one, and for a $\mathbb{Z}_2$ charge with
$\gamma=\pi N_{\mathcal C}$ the doubled phase $2\gamma = 2\pi N_{\mathcal C}$ is a
multiple of $2\pi$, hence indistinguishable from zero.
We instead exploit the chirality $S = \sigma_y \otimes (t\to -t)$ of the
Floquet Hamiltonian, $S\mathcal{H}S^{-1}=-\mathcal{H}$, which pairs Floquet modes
$|\phi_\pm\rrangle$ of opposite quasienergy, $q_+=-q_-$. 

\begin{figure}[b]
\includegraphics{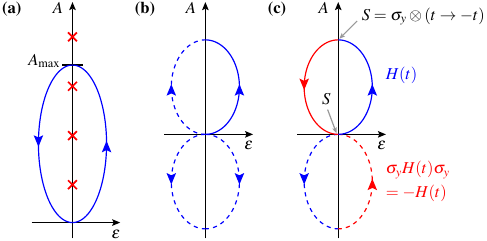}
\caption{(a) Loop in parameter space with Floquet--Berry phase $\gamma =
n\pi$, with $n$ the number of diabolic points enclosed by the loop.
(b) Symmetry-related semicircles.  For equal traversal times, on
all four arcs a given initial state acquires the same dynamical phase.
(c) Protocol for cancelling the dynamical phase.  The solid lines mark the
driving parameters, while dashed lines refer to the driving parameters of
the Hamiltonian related by chirality.}
\label{fig:protocol}
\end{figure}

\begin{figure}
\includegraphics{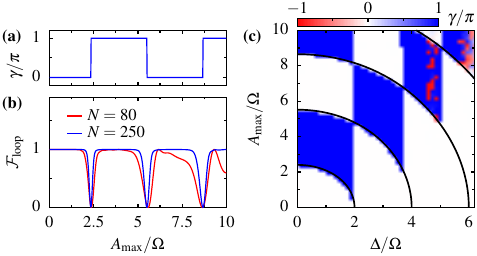}
\caption{(a) Floquet--Berry phase obtained with the chirality-based protocol sketched in Fig.~\ref{fig:protocol}(c)
for tunneling $\Delta=\Omega/2$, protocol time $T^*=250T$, and loop
parameter $\epsilon_\text{max}=0.2\Omega$.
(b) Return fidelity for protocol durations $T^*=80T$ (red) and $250T$ (blue).
(c) Floquet--Berry phase as a function of the tunneling $\Delta$ and the turning point
$A_\text{max}$. The solid lines mark the location of the exact crossing shown in
Fig.~\ref{fig:spectrum}(b).  For large $A_\text{max}$ and $\Delta$, the
much larger protocol time $T^*=5\times 10^{4}T$ has been used.}
\label{fig:FBphase}
\end{figure}

We focus on the elliptical loop sketched in Fig.~\ref{fig:protocol}(a) for
which the detuning and the amplitude are varied as
\begin{align}
\epsilon(t) ={}& \epsilon_\text{max}\sin\omega t ,
\label{eq:loop1}
\\
A(t) ={}& A_\text{max}(1-\cos\omega t)/2,
\label{eq:loop2}
\end{align}
with $\omega = \Omega/N$, where large $N$ will ensure adiabaticity.
Moreover, as we will see below, our protocol requires that $N$ is an even
integer.  At $t=0$, the loop starts at the origin of parameter space,
$\epsilon = A = 0$.  Between $A=0$ and the turning point $A_\text{max}$,
the loop encloses $n$ degeneracy points, such that we expect to find
$\gamma = n\pi \pmod{2\pi}$.  The key idea is to divide the loop into two
symmetry-related halves with opposite dynamical phase and equal
contribution to the Floquet--Berry phase.

Let us first consider the Floquet Hamiltonian as a function of the driving
parameters, $\mathcal{H}(\epsilon,A)$.  By a combination of $\sigma_x$ and
the time shift $t\to t+T/2$, one can invert the sign of the detuning and
the amplitude separately.  Since these transformations are unitary in Sambe
space, $\mathcal{H}(\pm\epsilon,\pm A)$ and $\mathcal{H}(\epsilon,A)$ have
the same quasienergies.  Therefore, Floquet modes that are adiabatically
connected to the ground state $|g\rangle$ at $\epsilon=A=0$ via one of the
four oriented semicircles in Fig.~\ref{fig:protocol}(b) give rise to the
same dynamical phases.  Their contribution to the Floquet--Berry phase, by
contrast, may have opposite sign, depending on the orientation of the arc.

After propagating the ground state along the solid line in
Fig.~\ref{fig:protocol}(b), one may continue with closing the loop via the
arc in the upper-left quadrant.  Then, however, the dynamical phases of
each arc add up and will dominate the total phase.  To obtain only the
Floquet--Berry phase, one may propagate the system during the second half
of the loop with the negative Hamiltonian.  In practice, one exploits the
chirality $\sigma_y H(t)\sigma_y = -H(t)$ by applying a $\sigma_y$-pulse at
the beginning and at the end of the second half loop.  For the usual following to
eigenstates, such pulses transfer the adiabatic ground state to the excited
state and back.  Here the situation is slightly more complicated, because
we deal with adiabatic following to Floquet modes.  Therefore, we have to
consider the chirality of the Floquet Hamiltonian, $S\mathcal{H}S^{-1} =
-\mathcal{H}$ with $S = \sigma_y\otimes (t\to-t)$, rather than that of the
Hamiltonian.  $S$ relates Floquet modes with opposite quasienergy according
to $\sigma_y|\phi_\mp(t)\rangle \sim |\phi_\pm(-t)\rangle$.  While time
cannot be inverted, the $T$-periodicity of the Floquet modes helps: for $t
= k T$, the relation becomes $\sigma_y|\phi_\mp(kT)\rangle \sim
|\phi_\pm(-kT)\rangle = |\phi_\pm(kT)\rangle$.  Therefore, at stroboscopic
times, the mapping $t\to -t$ in $S$ is without any effect.  This explains
the need for $T^*$ being a multiple of $T$.

Concretely, we consider the loop sketched in Fig.~\ref{fig:protocol}(c).
The protocol starts at the origin $A=\epsilon=0$ of
parameter space with a preparation in the ground state, which for zero
amplitude coincides with the Floquet mode $|\phi_-(0)\rangle$.  Then during
$N/2$ driving periods with $N$ even, the detuning and the amplitude are
adiabatically changed along the blue arc.  At time $T^*/2 = NT/2$, a
$\sigma_y$-pulse is applied to turn the Floquet state
$|\phi_-(T^*/2)\rangle$ into $|\phi_+(T^*/2)\rangle$.  Then the evolution
continues for another $T^*/2$ along the red arc, which corresponds to the
evolution of $|\phi_-\rangle$ with $-H(t)$. 
The geometric contributions accumulated on the two adiabatic arcs,
together with the endpoint phases associated with the two branch-exchange
pulses, combine into the gauge-invariant Floquet--Berry phase of the closed
loop, as is demonstrated explicitly in Appendix~\ref{app:symmetry}.
Finally, a second $\sigma_y$-pulse is applied. The resulting state then
reads $|\psi_{\rm final}\rangle = e^{i\gamma}|\psi_{\rm initial}\rangle$,
with $\gamma$ a multiple of $\pi$ equal to the number of enclosed
degeneracies with nontrivial topological charge.

As $A_\text{max}$ increases, $\gamma$ jumps between $0$ and $\pi$ each time
an additional topological degeneracy enters the loop, which can be
appreciated in Fig.~\ref{fig:FBphase}(a).  A breakdown of adiabaticity,
signaled by the fidelity $\mathcal{F}_\text{loop} =
|\langle\psi_\text{final}|\psi_\text{init}\rangle|^2$, occurs for large
$A_\text{max}$, in particular when the protocol time $T^*$ is relatively
short, see Fig.~\ref{fig:FBphase}(b).  For sufficiently slow parameter
variation, it occurs only in narrow windows where the loop comes close to a
cone tip. Away from these windows, the protocol cleanly returns the initial
state.

The result as a function of $A_\text{max}$ and also the tunnel matrix
element $\Delta$ [Fig.~\ref{fig:FBphase}(c)] reproduces the structure of
the location of the crossings shown in Fig.~\ref{fig:spectrum}(b).
Remarkably, the Floquet--Berry phase abruptly changes between $0$ and $\pi$
when for increasing $\Delta$, a crossing moves to the origin and
disappears. This underlines the topological character of the Floquet--Berry
phase. For large tunneling and large $A_\text{max}$, it becomes
increasingly difficult to achieve adiabaticity, as is visible in the
deviations from $\gamma=\pi$ in the upper-right corner.

\begin{figure}
\includegraphics{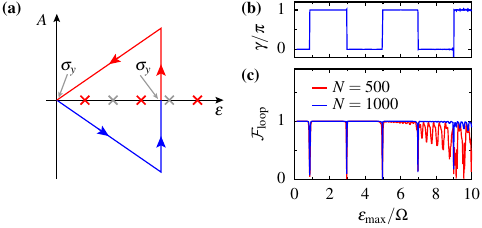}
\caption{(a) Protocol for probing the topological charges at zero
amplitude.  Only the degeneracies with odd index $m$ (red crosses) are
topologically non-trivial, cf.\ Fig.~\ref{fig:bargmann}(a).
(b) Floquet--Berry phase as a function of the maximal detuning for tunneling
$\Delta=\Omega/2$ and protocol time $T^*=1000T$.
(c) Corresponding return fidelity for $T^*=500T$ (red) and $1000T$
(blue).}
\label{fig:FBphaseA0}
\end{figure}

The loop in Fig.~\ref{fig:protocol} can be modified readily such that
the protocol becomes sensitive to the topological charges of the $m$-photon
resonances at zero amplitude.  Here one can exploit the fact that the
quasienergies do not depend on the sign of $A$. Therefore, cancellation of
the dynamical phase can be achieved by a loop with reflection symmetry at
the $\epsilon$-axis together with applying $\sigma_y$-pulses.
To facilitate adiabatic following, the loop should
stay away from the degeneracies at finite amplitude and should avoid the
regions with tiny splitting, see Fig.~\ref{fig:bargmann}(b).  A suitable
choice is the triangle sketched in Fig.~\ref{fig:FBphaseA0}(a). To facilitate the
preparation of the initial state, it also starts at the origin. It
continues along a bisection line until the detuning reaches a value
$\epsilon_\text{max}$.  Then the driving is adiabatically switched off.
After the application of a $\sigma_y$-pulse, the loop is closed on the
other half-plane and terminates with a further $\sigma_y$-pulse.

The result of a numerical propagation is shown in
Figs.~\ref{fig:FBphaseA0}(b,c).  The Floquet--Berry phase again becomes
$0$ and $\pi$, depending on the number of enclosed degeneracies with a
topological charge.  In particular, this verifies that only the $m$-photon
resonances with odd $m$ have a topological charge, while those with even
$m$ are topologically trivial.
Figure~\ref{fig:FBphaseA0}(c) indicates that now adiabatic following is
harder to achieve.  For the resonances of higher order, even the relatively
large protocol time $T^*=500T$ leads to a significant reduction of the loop
fidelity $\mathcal{F}_\text{loop} =
|\langle\psi_\text{final}|\psi_\text{initial}\rangle|^2$.

The requirements of the schemes appear compatible with present-day platforms.
It requires only adiabatic following of a single Floquet mode over a
coherence time $T^*=NT$, together with two fast $\sigma_y$ pulses applied
at $T^*/2$ and $T^*$.  Taking $N\sim10^3$ driving periods keeps the
evolution adiabatic everywhere except in the immediate neighborhood of a
cone tip, which the loop should be designed to avoid; the fidelities in
Figs.~\ref{fig:FBphase}(b) and \ref{fig:FBphaseA0} confirm that clean returns are then obtained.
On the same shaken-lattice platform, the time-reversal and parity
symmetries that protect the charge can be tuned with precision
\cite{HebraudNJP26}, providing a direct experimental handle on the
symmetry-breaking analysis of Sec.~\ref{sec:topology}.

\section{Measurement via an ancilla}
\label{sec:measurement}

Any interferometric measurement of a phase requires a phase reference.  For
a two-level system, this can be achieved by preparing a superposition of
ground and excited states of $H(0)$, whose Berry phases have opposite sign.
In the present case, this is insufficient, because under the
chirality-based protocol, the two states acquire opposite phases $\pm\pi$,
so that their relative phase becomes $2\gamma = 0$.  We address this by a
Ramsey scheme \cite{RamseyPR50} for which we introduce an ancilla level
$|a\rangle$ whose energy $\delta(t)$ is controllable and not coupled
off-diagonally to the two-level system except by external $\pi/2$-pulses.

\begin{figure}
\includegraphics{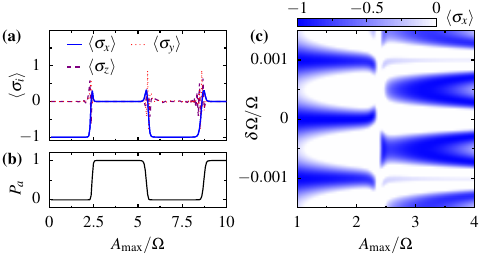}
\caption{Protocol with ancilla state for tunnel coupling $\Delta=0.3\Omega$ and
protocol duration $T^*=1000T$.
(a) Components of the Bloch vector of the final state projected to the
two-level subspace as a function of the turning point $A_\text{max}$.
(b) Corresponding ancilla population.
(c) $x$-component of the Bloch vector shown in panel (a) for
$E_a=2.5\Omega$ and slightly varied frequency $\Omega+\delta\Omega$.  The
protocol time $T^*$ has been adjusted, while all other parameters are
kept equal.}
\label{fig:ancilla}
\end{figure}

The protocol then runs as follows. After preparation in the ground state
$|g\rangle= (0,1,-1)^T/\sqrt{2}$ (the two-level ground state, with
vanishing amplitude on the ancilla), a $\pi/2$-pulse $U_\text{pulse} =
\exp(i\pi G/4)$ with $G = |a\rangle\langle g| + |g\rangle\langle a|$
creates the superposition $(|g\rangle+|a\rangle)/\sqrt{2}$.  Then under the
chirality protocol of Sec.~\ref{sec:protocol}, the ground state acquires
the Floquet--Berry phase, while the phase of the ancilla, $\varphi_a =
-\int_0^{T^*} dt'\delta(t')$ is purely dynamical. Reversing the
$\pi/2$-pulses and measuring the populations of $|g\rangle$, $|e\rangle$,
and $|a\rangle$ yields fringes that depend on $\gamma-\varphi_a$. The
ancilla population follows the Ramsey law \cite{RamseyRMP90}
\begin{equation}
P_a = \tfrac{1}{2}\big[\,1 - \cos(\gamma +\varphi_{\rm pulse} -\varphi_a)\,\big] ,
\label{eq:ramsey}
\end{equation}
where $\varphi_{\rm pulse}=\pi$ is the known offset produced by the two
$U_\text{pulse}$ pulses; it is a fixed calibratable constant and simply
relabels the fringes.  In particular, for $\gamma+\varphi_{\rm pulse}
-\varphi_a = 0$, the wave function returns to its original value, while for
$\gamma +\varphi_{\rm pulse} -\varphi_a = \pi$, the system ends up in the
ancilla state.  To obtain such a clear signal, the ancilla phase must be
controlled, ideally such that it becomes a multiple of $2\pi$.  The
simulation shown in Figs.~\ref{fig:ancilla}(a,b) confirms this expectation.
When an even number of crossings is enclosed by the loop,
$\langle\sigma_x\rangle=-1$ which corresponds to the ground state
$|g\rangle$. For an odd number, the Bloch vector vanishes, and the ancilla
is fully populated.

In practice, it may be difficult to control the ancilla phase with
sufficient precision, such that $\varphi_a$ remains unknown. Nevertheless,
a clear signal can be obtained under the weaker condition that $E_a$ is
unknown but can nevertheless be kept constant.  The idea is to repeat the
experiment with tiny variations of the driving frequency
$\Omega\to\Omega+\delta\Omega$, such that the ancilla phase becomes
$\varphi_a = 2\pi N E_a/(\Omega + \delta\Omega)$.  Since $N\gg1$, even for
small $\delta\Omega$ the effect may be significant.  By contrast, its
impact on the position of the crossing is negligible.  Then, as a function
of $A_\text{max}$ and $\delta\Omega$, $\langle\sigma_x\rangle$ assumes the
pattern depicted in Fig.~\ref{fig:ancilla}(c).  It shows that upon slight
variation of $\delta\Omega$, the Ramsey fringes switch between constructive
and destructive interference.  When at $A_\text{max} \approx 2.4\Omega$,
the Floquet--Berry phase changes from $0$ to $\pi$, the pattern becomes
inverted. This represents a distinct fingerprint of the topological charge
as soon as the loop is extended beyond a crossing.

The scheme is conceptually analogous to the measurement of the Zak phase in
optical lattices \cite{AtalaNP13}, and to the Aharonov--Bohm
interferometric determination of Bloch-band topology \cite{DucaS15}, where
the dynamical phase contribution is removed by a spin-echo sequence to an
internal degree of freedom. Its implementation should be within reach of
superconducting qubits with a tunable third level \cite{LeekS07}, or of
cold atoms in optical lattices where the two driven levels are dressed
momentum states and the ancilla is an internal hyperfine sublevel.

\section{Conclusion} \label{sec:conclusion}

We have shown that the exact quasienergy crossings of a strongly driven
two-level system carry a quantized $\mathbb{Z}_2$ topological charge in the
parameter plane of detuning and amplitude, measured by a Floquet--Berry phase
$\gamma(\mathcal{C}) = \pi N_{\mathcal C}\ (\mathrm{mod}\ 2\pi)$ for a loop
$\mathcal{C}$ enclosing $N_{\mathcal C}$ crossings. The conical intersections at integer detunings
$\epsilon=n\Omega$ each carry the nontrivial charge $\pi$, in
accordance with their generic diabolic structure.  In turn, the persistence
of the quantized phase as the contour is shrunk toward a crossing provides a
diagnostic that the degeneracy is exact. The
degeneracies at $m$-photon resonances and zero amplitude, by contrast, are
non-diabolic for $m>1$: along the Stark-shifted resonance line the minimum
gap opens as $|A|^m$, and they obey a parity
selection rule, carrying the nontrivial $\mathbb{Z}_2$ charge for odd $m$ and a
trivial charge for even $m$. This parity originates in the order $m$ at which the
signed driving amplitude enters the off-diagonal coupling in Sambe-space
perturbation theory, the coupling changing sign across $A=0$ only for odd $m$.

Our chirality-based protocols cancel the dynamical phase by exploiting an
anti-symmetry of the Floquet Hamiltonian itself, $S\mathcal{H}S^{-1} =
-\mathcal{H}$, rather than that of
the bare Hamiltonian. This is essential in the strong-driving regime,
where the standard spin-echo would identify topologically distinct
configurations as equivalent. Combined with an ancilla-based readout, the
protocol provides a route to direct interferometric measurement of the
topological charge.

Several extensions come to mind. Since the chirality argument depends on
the spatio-temporal structure of the drive only through $S\mathcal{H}S^{-1}
= -\mathcal{H}$, it may be generalized to other models with the same
symmetry, including arrays of driven qubits where braiding of topological
charges may become accessible \cite{SchindlerPRX25}. The dissipative response near the
non-diabolic degeneracies, based on the analysis of
Ref.~\cite{KohlerPRA24}, should be sensitive to the parity selection rule
in a measurable way. Finally, the parity rule itself is a small instance
of a more general question: how do degeneracies of order $m$ in one
parameter direction and order $1$ in the other contribute to the
topology of multi-band systems, in any setting where such anisotropic
band touchings arise.

\begin{acknowledgments}
This work was supported by the Institut Universitaire de France and by the
Spanish Ministry of Science, Innovation, and Universities under Grant Nos.\
PID2023-149072NB-I00 and AIA2025-163435-C44.
\end{acknowledgments}

\appendix
\section{Perturbation theory at the degeneracy points}
\label{app:perturbation}

We have argued that owing to time-reversal symmetry of our system, the Berry curvature must vanish
except at isolated degeneracy points.  Therefore, it is sufficient to focus on an arbitrarily small
neighborhood of a degeneracy point at which we denote detuning,
amplitude, and Floquet Hamiltonian by $\epsilon_0$, $A_0$, and $\mathcal{H}_0$, respectively.
In its vicinity, the system is described by $\mathcal{H}_0$ and the perturbation
$\mathcal{H}_1 = V+W$ with
\begin{gather}
V = \frac{1}{2}(\epsilon-\epsilon_0) \sigma_z,
\\
W = \frac{1}{2}(A-A_0)\sigma_z \cos(\Omega t).
\end{gather}
Perturbation theory will reveal the nature of the intersection.

\subsection{Integer detuning}

For integer detuning $\epsilon = n\Omega$, a hidden symmetry allows one to
classify the Floquet states as even and odd, where quasienergies with
different parity may form exact crossings \cite{KohlerQ26}.  Their location
in parameter space can be estimated by the common Bessel function
approximation \cite{StrassPRL05,AshhabPRA07,IvakhnenkoPR23} and beyond
\cite{CreffieldPRB03,DellaVallePRL07,WeiJCP26}.  Following
Ref.~\cite{KohlerPRA24}, we choose two degenerate Floquet states of
$\mathcal{H}_0$ with different parity as basis and evaluate the matrix
elements of $V$ and $W$ in Sambe space.

In the absence of detuning, $\epsilon=0$, the symmetry is explicit and known as
generalized parity $G = \sigma_x\otimes(t\to t+T/2)$ \cite{PeresPRL91}.  Under mapping with
$G$, the perturbations $V$ and $W$ are odd and even, respectively.  Hence,
evaluating their matrix elements yields \cite{KohlerPRA24}
\begin{equation}
\mathcal{H}_\text{eff} = \epsilon v_x\tau_x + (A-A_0)w_z\tau_z \,.
\label{eq:H1integer}
\end{equation}
Because $V$ is odd under the generalized parity $G$, its diagonal
matrix elements vanish and its generically nonzero matrix element is the
off-diagonal one. Conversely, $W$ is even under $G$, so its off-diagonal
matrix element vanishes and the difference of its diagonal elements produces
the $\tau_z$ component. With the convention
$\tau_z=\mathrm{diag}(1,-1)$ in the basis
$(|\varphi_+\rrangle,|\varphi_-\rrangle)$, the coefficients are
\begin{gather}
v_x = \frac{1}{2}\llangle\varphi_+|\sigma_z|\varphi_-\rrangle ,
\label{eq:vx-corrected}\\
w_z = \frac{1}{4}\big[
\llangle\varphi_+|\sigma_z\cos(\Omega t)|\varphi_+\rrangle
-
\llangle\varphi_-|\sigma_z\cos(\Omega t)|\varphi_-\rrangle
\big] .
\label{eq:wz-corrected}
\end{gather}
Both coefficients are generically nonzero.
The resulting spectrum is a deformed Dirac cone
with an elliptical shape and primary axes parallel to the parameter axes.

For the crossings at non-zero integer detuning, the hidden symmetry allows
one to still classify the Floquet modes as even or odd.  However, its
individual terms such as $V$ and $W$ may not possess the same symmetry,
which implies that also coefficients $v_z$ and $w_x$ emerge.  The spectrum
is again a deformed Dirac cone, but now with an arbitrary orientation.

\subsection{Multi-photon resonances}

For zero amplitude, the driving enters only formally. Then the eigenstates
$|g\rangle$ and $|e\rangle$ of the undriven Hamiltonian $H_0 =
\Delta\sigma_x/2 + \epsilon_0\sigma_z/2$ can be considered as Floquet modes
with quasienergies equal to the eigenenergies, $q_\pm = E_{e,g}$.  A phase
factor $e^{in\Omega t}$ with integer $n$ maps them to equivalent Floquet
modes with quasienergies $q_\pm^{(n)} = E_{e,g}+n\Omega$, as can be
demonstrated by inserting into Eq.~\eqref{eq:Feq}.  Here, we focus
on a $m$-photon resonance $E_e = E_g+m\Omega$, for which the Floquet modes
$|\varphi_-(t)\rangle = |g\rangle$ and $|\varphi_+(t)\rangle = e^{-im\Omega
t}|e\rangle$ are degenerate.

For $m=1$, we proceed as above and evaluate the matrix elements of $V$ and
$W$.  The particular form of $|\varphi_\pm(t)\rangle$ separates the
expressions into a matrix element of the undriven system and a time
integral.  This provides for the perturbation the expression
\begin{equation}
\mathcal{H}_1 = \frac{1}{2}(\epsilon-\epsilon_0)\langle e|\sigma_z|e\rangle\tau_z
+ \frac{1}{4}A \langle g|\sigma_z|e\rangle \tau_x .
\label{eq:m1}
\end{equation}
It is the same as Eq.~\eqref{eq:H1integer}, but with detuning and amplitude
interchanged.  Thus, the Floquet spectrum also forms a Dirac cone.

\begin{figure}
\centerline{\includegraphics{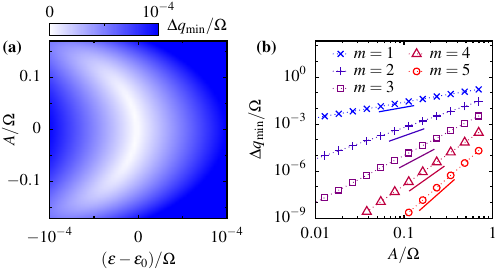}}
\caption{Quasienergy splitting in the vicinity
of the $m$-photon resonances for tunneling $\Delta=\Omega/2$.
(a) Minimal splitting for $m=3$.
(b) Scaling of the minimum as a function of the amplitude.
The solid lines indicate the power law $\Delta q_\text{min}\propto |A|^m$.}
\label{fig:scaling}
\end{figure}

For $m>1$, the operator $V$ again provides the first term in
Eq.~\eqref{eq:m1}, i.e., it is diagonal in the new basis.  The contribution
of $W$, however, vanishes in first order due to the time integration.
Hence, $\mathcal{H}_\text{eff} = (\epsilon-\epsilon_0)\langle
e|\sigma_z|e\rangle\tau_z/2$, which does not capture the impact of the
amplitude variation and forces us to consider higher-order terms.
These terms cause an ac-Stark shift, which affects the level
splitting, such that the location of the minimal splitting shifts.  For the
three-photon resonance, this can be appreciated from the numerical data
shown in Fig.~\ref{fig:scaling}(a).

We introduce the resonance mismatch $\eta(\epsilon,A) = E_e - E_g - m\Omega
\simeq \alpha\,(\epsilon-\epsilon_0)$, which has to be determined self-consistently.
It contributes a term $\tfrac{1}{2}\eta\,\tau_z$.  The lowest non-vanishing contribution
in $A$ arises at order $W^m$ and is directed along $\tau_x$. Moreover,
time-reversal symmetry forbids a $\tau_y$ component, so the effective
two-level Hamiltonian in the degenerate subspace has the generic form
\begin{equation}
\mathcal{H}_{\rm eff}
= d_x(\eta,A)\,\tau_x + d_z(\eta,A)\,\tau_z ,
\label{eq:app-Wzero}
\end{equation}
where the absence of a parameter-dependent identity term can be established
without relying on the perturbative expansion by the following argument. Since the physical
Hamiltonian~\eqref{eq:H} is traceless at every time,
$\det U(T)=\exp[-i\int_0^Tdt\,\mathrm{Tr}\,H(t)]=1$. Thus the two
eigenvalues of the one-period propagator form the pair
$e^{\pm i\theta}$. At the $m$-photon resonance,
$E_e-E_g=m\Omega$ and $E_e=-E_g$, hence
$U(T)=(-1)^m\mathbb I$. In a neighborhood of the degeneracy, one may
therefore write
$U(T)=(-1)^m\exp[-i h_{\rm eff}T]$ with $h_{\rm eff}$ traceless.
Equivalently, after subtracting the constant zone-center or zone-edge
quasienergy $q_0=m\Omega/2\pmod{\Omega}$, the effective Floquet
Hamiltonian contains only Pauli components. No parameter-dependent
contribution proportional to $\mathbb I$ can occur.

The remaining coefficients are of the form
\begin{align}
d_x(\eta,A) &= \beta_m A^m + \mathcal{O}(\eta A^m,\,A^{m+2}) ,
\label{eq:app-dx}\\
d_z(\eta,A) &= \tfrac{\alpha}{2}\,\eta + c_2 A^2 + c_4 A^4 + \dots
+ \mathcal{O}(\eta^2) ,
\label{eq:app-dz}
\end{align}
and can be evaluated via a Sambe-space perturbation expansion. The coefficient \(\beta_m\) collects the amplitudes of the \(m\)-step virtual one-photon processes connecting the two resonant Sambe states; it is generically nonzero and vanishes only on nongeneric parameter manifolds.
The even
powers of $A$ in $d_z$ are ac-Stark shifts which displace the resonance
line locally, $\eta\to\eta+(2/\alpha)(c_2A^2+\dots)$.  Along the line with
$d_z=0$, which has the shape of the white area in
Fig.~\ref{fig:scaling}(a), the quasienergy splitting is fully determined by
$d_x$ and is expected to behave like $\Delta q\propto |A|^m$.  We have
numerically verified this behavior for $m=1,\ldots,5$, see
Fig.~\ref{fig:scaling}(b).

\section{Phases acquired on symmetry-related arcs}
\label{app:symmetry}

The protocol introduced in Sec.~\ref{sec:protocol} is based on the
cancellation of the dynamical phases acquired on the blue and the red
contour in Fig.~\ref{fig:protocol}(c). 
In addition, the geometric phases accumulated on the two
arcs, together with the phases generated by the two branch-exchange pulses,
must reproduce the Berry holonomy of one Floquet branch around the closed
loop. We demonstrate both statements without fixing the local $U(1)$
Floquet gauge.

\subsection{Dynamical phase}

The cancellation of the dynamical phase requires two symmetries.  First,
the chirality of the Floquet Hamiltonian provides for the quasienergies the
relation $q_-(\epsilon,A) = -q_+(\epsilon,A)$.  The second ingredient is
that the quasienergies do not depend on the sign of the detuning
$\epsilon$.  Formally, this can be concluded from mapping the Floquet
Hamiltonian with $C = \sigma_x\otimes(t\to t+T/2)$, for which
$C\mathcal{H}(\epsilon,A)C^{-1} = \mathcal{H}(-\epsilon,A)$.  Since $C$ is
a unitary transformation in Sambe space, it leaves the eigenvalues of
$\mathcal{H}$, i.e.\ the quasienergies invariant.

The dynamical phase during the second half of the cycle reads
\begin{equation}
\Phi_+^{(2)}
= -\int_{T^*/2}^{T^*} dt\, q_+(\epsilon(t),A(t)).
\end{equation}
On the contour in Eqs.~\eqref{eq:loop1} and \eqref{eq:loop2} the parameters
obey $\epsilon(t) = -\epsilon(T^*-t)$ and $A(t) = A(T^*-t)$, which we use
to substitute the integration variable, $t\to T^*-t$, to obtain
\begin{equation}
\Phi_+^{(2)} = -\int_0^{T^*/2} dt\, q_+(\epsilon(t),A(t)) = -\Phi_-^{(1)}.
\end{equation}
This implies cancellation of the dynamical phases.

One may ask whether this finding is consistent with the Brillouin zone
invariance of Floquet theory.  This invariance means that the Floquet modes
$|\phi(t)\rangle$ and $e^{ik\Omega t}|\phi(t)\rangle$ with the quasienergies $q$
and $q+k\Omega$ correspond to the same solution of the Schr\"odinger
equation, despite being different solutions of the Floquet
equation~\eqref{eq:Feq}. Therefore, any physical property must not depend
on the choice of one of infinitely many equivalent Floquet modes.  Here, this is
granted, because for our protocol time $T^*=NT$ with $N$ even, replacing a
Floquet mode by an equivalent one changes the dynamical phase merely by a
multiple of $2\pi$.

\subsection{Floquet--Berry phase}

The protocol in Sec.~\ref{sec:protocol} contains two
$\sigma_y$ pulses which exchange the two Floquet branches at stroboscopic
times. The appropriate statement is most transparent without choosing a
special gauge. Chirality,
\begin{equation}
S\mathcal H S^{-1}=-\mathcal H,
\quad
S=\sigma_y\otimes(t\to-t),
\quad S^2=\mathbb I ,
\end{equation}
maps the two branches onto each other. In a general local $U(1)$ Floquet
gauge this relation contains a parameter-dependent phase,
\begin{equation}
\begin{split}
S|\phi_-(\bm x)\rrangle &= e^{i\vartheta(\bm x)}
|\phi_+(\bm x)\rrangle, \\
S|\phi_+(\bm x)\rrangle &= e^{-i\vartheta(\bm x)}
|\phi_-(\bm x)\rrangle .
\end{split}
\label{eq:chirality-gauge}
\end{equation}
Defining the two Berry connections by
\begin{equation}
\mathcal A_\mu^\pm(\bm x)
=\llangle\phi_\pm(\bm x)|i\partial_\mu|
\phi_\pm(\bm x)\rrangle ,
\end{equation}
and using the fact that $S$ is independent of the slow parameters gives
\begin{equation}
\mathcal A_\mu^+(\bm x)
=\mathcal A_\mu^-(\bm x)+\partial_\mu\vartheta(\bm x).
\label{eq:connection-gauge}
\end{equation}
Thus the equality of the two connections holds in the
symmetry-adapted gauge $\vartheta=0$, but it is not itself a
gauge-invariant statement.

Let $\mathcal C_1$ denote the first half of the loop, from the initial
point $\bm x_0$ to the switching point $\bm x_m$, and $\mathcal C_2$ the
second half, from $\bm x_m$ back to $\bm x_0$. The system follows the
minus branch on $\mathcal C_1$ and the plus branch on $\mathcal C_2$.
At the stroboscopic switching times, the time-reflection part of $S$ acts
trivially on the $T$-periodic Floquet mode, so the two $\sigma_y$ pulses
implement Eq.~\eqref{eq:chirality-gauge}. Consequently, they contribute
the endpoint phases $+\vartheta(\bm x_m)$ and
$-\vartheta(\bm x_0)$, respectively. The complete geometric phase of
the piecewise protocol is therefore
\begin{equation}
\Gamma_{\rm geom}
=\int_{\mathcal C_1}\mathcal A_\mu^-\,dx^\mu
+\int_{\mathcal C_2}\mathcal A_\mu^+\,dx^\mu
+\vartheta(\bm x_m)-\vartheta(\bm x_0).
\label{eq:geom-piecewise}
\end{equation}
Using Eq.~\eqref{eq:connection-gauge} and the orientation of
$\mathcal C_2$,
\begin{align}
\Gamma_{\rm geom}
 ={}&\int_{\mathcal C_1}\mathcal A_\mu^-\,dx^\mu
 +\int_{\mathcal C_2}\mathcal A_\mu^-\,dx^\mu
\nonumber\\
& +\vartheta(\bm x_0)-\vartheta(\bm x_m)
  +\vartheta(\bm x_m)-\vartheta(\bm x_0)\nonumber\\
={}&\oint_{\mathcal C}\mathcal A_\mu^-\,dx^\mu
=\gamma(\mathcal C)\pmod{2\pi}.
\label{eq:geom-gauge-invariant}
\end{align}
The gauge-dependent endpoint terms therefore cancel exactly between the
adiabatic arcs and the branch-exchange pulses. The physical protocol
measures the Berry holonomy of the closed loop independently of the local
$U(1)$ gauge chosen for either Floquet branch. The discrete freedom to
shift a Floquet mode by $e^{ik\Omega t}$ is the Brillouin-zone freedom
discussed in the preceding subsection; for $T^*=NT$ with even $N$ it
changes the accumulated phase only by multiples of $2\pi$.

\end{document}